\documentclass[aps, amsmath,amssymb, prd,twocolumn,showpacs,superscriptaddress,groupedaddress,nofootinbib]{revtex4-1}

\usepackage{graphicx}
\usepackage{dcolumn}
\usepackage{bm} 
\usepackage{xcolor}    

\usepackage{color}
\usepackage{soul}
\usepackage{changes}

\usepackage{sidecap}
\usepackage{wrapfig}       
\usepackage{subfig}  
\usepackage{float}

\begin{document} 

\title{$k$-Strings with exact Casimir law and Abelian-like profiles} 

\author{L. E. Oxman}
 \email{oxman@if.uff.br}
\author{G. M. Sim\~oes} 
 \email{gustavoms@fisica.if.uff.br}
\affiliation{
Instituto de F\'isica, Universidade Federal Fluminense,
Campus da Praia Vermelha, Niter´oi, 24210-340, RJ, Brazil.
}

\begin{abstract} 
We explore vortex solutions for a class of dual $SU(N)$ Yang-Mills models with $N^2-1$ Higgs fields in the adjoint representation. 
Initially, we show that there is a collective
behavior that can be expressed in terms of a small $N$-independent number of field profiles. Then, we find a region in parameter space 
where the nontrivial
profiles coincide with those of the Nielsen-Olesen vortex, and the energy scales exactly with the quadratic Casimir. Out of this region, we solve the ansatz equations numerically and find very small deviations from the Casimir law. The coexistence of Abelian-like string profiles and non-Abelian scaling features is welcome, as these properties have been approximately observed in pure YM lattice simulations. 

\end{abstract}

\maketitle

\section{\label{Intro}Introduction} 

Although dual superconductor models have long been proposed as an effective description of color confinement 
\cite{Nambu1974,Mandelstam1976,'tHooft19781}, so far, no such model 
has been completely successful. The many candidates 
\cite{Maedan:1989ju, Koma:2001ut, Koma:2003hv, Baker:1989qp, Baker:1991mu, Baker:1998jw, Hanany:2003hp, Hanany:2004ea, Tong:2003pz, 
Tong:2008qd, Auzzi:2003fs, Konishi:2007dn, Gorsky:2004ad, Armoni:2003nz, Shifman:2004dr, Konishi:2001cj, Kneipp:2001tp, 
Kneipp:2007fg,AuzziKumar, Oxman2013}
grasp some, but not all, of the rich confinement phenomenology obtained from the lattice. This includes the 
static quark-antiquark potential \cite{Bali:2000gf}, the L\"uscher term \cite{Luscher:2002qv}, the 
Abelian-like transverse chromoelectric field profile \cite{Cosmai2017}, and Casimir scaling \cite{Bali2000CS}. 
The latter refers to the dependence of the string tension with the quadratic Casimir operator of the quark representation, at intermediate distances.
Another important feature to be accommodated is the asymptotic string tension scaling-law, which can only depend on an integer $k$ (modulo $N$) that dictates how the center of $SU(N)$ is realized in the quark representation ($N$-ality). In accordance with Monte Carlo simulations in four dimensions \cite{Teper2004}, the Sine and Casimir laws are among the possible behaviors. Note that the latter corresponds to scaling with the quadratic Casimir of the $k$-antisymmetric representation. 
$N$-ality suggests that confining strings 
could be represented as stable topological vortices in a Yang-Mills-Higgs (YMH) field description. Models with 
fields transforming in the fundamental representation
\cite{Gorsky:2004ad}, the adjoint \cite{Konishi:2001cj, Kneipp:2001tp, 
Kneipp:2007fg, AuzziKumar, Oxman2013}, or both \cite{Hanany:2003hp, Hanany:2004ea, Tong:2003pz, 
Tong:2008qd, Auzzi:2003fs, Konishi:2007dn, Shifman:2004dr}, are among 
the possibilities. In spite of the fact that these models possess vacua leading to confining strings 
with $N$-ality,  the different field contents and Higgs potentials make it necessary to work on a case by case basis to 
determine the precise vortex profiles and the behavior of the string tension. 
For example, a model motivated by supersymmetry and based on three complex adjoint fields was analyzed in Ref. \cite{AuzziKumar}. Although
the group action on the vacua manifold is not transitive in this case,
the physical properties in the different sectors can be related by means of appropriate mappings between them.
Moreover, a numerical analysis of the vortex solutions showed a string tension closely approximated by a Casimir law.

Another important (more direct) approach to describe confinement, 
developed over the years, is based on the detection of ensembles of magnetic defects that could capture the path-integral measure in lattice pure Yang-Mills theory. Quantum variables such as center-vortex 
worldsurfaces and monopole worldlines are among the most promising detected defects \cite{PhysRevD.55.2298, LANGFELD1998317, PhysRevD.58.094501, PhysRevLett.82.4582, 1126-6708-2000-02-033, PhysRevD.61.054504, 
ENGELHARDT2000249, PhysRevD.64.074504, REINHARDT2002133, GATTNAR2005105}. In particular, 
 Casimir scaling at intermediate distances can be understood as due to the finite thickness of center-vortex variables 
\cite{Greensite1998}. At asymptotic distances, these defects also implement $N$-ality, but their thickness cannot affect the string tension.  In this regime, the linear $k$-scaling, expected to occur in the large $N$ limit,
was reproduced by including monopole variables \cite{GreensiteOlejnik2002}.
Recently, we showed that an ensemble of two-dimensional percolating worldsurfaces with attached monopole worldlines in 4d 
can be related to a YMH effective model. In the effective description, the dual gauge field
represents the Goldstone modes in a condensate of one-dimensional defects, which generate the worldsurfaces, while a set of adjoint Higgs fields reproduce the monopole degrees of freedom.
The field content in Ref. \cite{Oxman:2018dzp} was chosen so as to implement the monopole fusion rules; in particular, 
models with an adjoint flavor index naturally encompass all possibilities. In this case, the phenomenological parameters can be chosen so as to obtain a transitive group action and drive $SU(N) \to Z(N)$ SSB. Transitivity of the vacua manifold automatically renders the different choices (labeled by points in $SU(N)/Z(N)$) physically equivalent.  
Then, among the alternatives, a detailed analysis of this type of model is of special interest. In this work,
despite the large number of fields, we will show that the system acquires a collective behavior where the classical vortex solutions are well accommodated by a small ($N$ and $k$-independent) number of profiles. Moreover, we shall obtain a region
in parameter space where the exact Casimir law holds. In this regime, most of the field profiles become frozen at their vacuum value while the nontrivial ones obey the Nielsen-Olesen equations, thus reproducing the chromoelectric field measured in the lattice 
(see sections \ref{Review} and \ref{Core}). 
In \ref{Numerics}, we will show the result of numerical simulations in other regions. Finally, in section \ref{Conclusions}, we will present our conclusions.

\section{\label{Review} The effective YMH model} 

A wide class of $SU(N)$ Yang-Mills-Higgs models can be given by the general action
\begin{equation}
\label{action}
S=\int d^4x  \frac{1}{4} \langle F_{\mu\nu},F^{\mu\nu}\rangle + \frac{1}{2}\langle D_\mu \psi_A, D^\mu \psi_A \rangle - V_{Higgs}(\psi_A)\;,
\end{equation}
where $F_{\mu\nu}$ is the non-Abelian dual field strength tensor,
\begin{equation}
    F_{\mu\nu} = \left[D_\mu,D_\nu\right] \makebox[.5in]{,} \quad D_\mu = \partial_\mu + ig [A_\mu, ]\;. 
\end{equation}
 The Killing form $\langle\,,\rangle$ is defined in the Lie algebra as
\begin{equation} 
\langle X,Y\rangle = {\rm Tr}({\rm Ad}(X) \, {\rm Ad}(Y))\;,
\end{equation}
where $Ad()$ stands for the adjoint representation. 
In Ref. \cite{Oxman2013}, the flavor index $A$ was chosen to run from $1$ to $N^2-1$, so that the number of Higgs fields 
matches the dimension of the $\mathfrak{su}(N)$ Lie algebra. With this matching, if the manifold of vacuum configurations 
$\mathcal{M}$ is given by $SU(N)$-rotated generators $\psi_A= vS T_A S^{-1}$, $[T_A, T_B]= f_{ABC} T_C$, 
then $N$-ality is naturally implemented via the 
spontaneous symmetry breaking pattern $SU(N)\rightarrow Z(N)$. The quartic potential
\begin{equation}
    \langle \psi_A\wedge\psi_B-vf_{ABC}\psi_C\rangle^2
\end{equation}
would lead to these vacua, however, it would also lead to a degenerate trivial vacuum $\psi_A =0$. Then, we 
expanded this expression and introduced independent coefficients for each term, thus proposing a natural potential
\begin{eqnarray}
\label{higgs}
V_{\rm Higgs}(\psi_I) &=& c+\frac{\mu^2}{2}\langle \psi_A,\psi_A\rangle 
+ \frac{\kappa}{3}f_{ABC}\langle \psi_A \wedge \psi_B, \psi_C\rangle \notag\\
&&+ \frac{\lambda}{4}\langle \psi_A\wedge \psi_B,\psi_A\wedge \psi_B\rangle\;,
\end{eqnarray}
to construct the effective field model ($c$ is adjusted such that $V_{Higgs}=0$ on $\mathcal{M}$). In this manner,
we obtained a region in parameter space that only leads to nontrivial vacua, characterized by
\begin{equation}
    v = -\frac{\kappa}{2\lambda}+\sqrt{\left(\frac{\kappa}{2\lambda}\right)^2-\frac{\mu^2}{\lambda}}\;.
\end{equation}

\section{\label{Core}The Vortex ansatz} 
As usual, in order to represent a straight infinite vortex along the z-axis, we set
\begin{subequations}
\label{ansatzbegin}
\begin{align}
    A_i &= S\mathcal{A}_iS^{-1} + \frac{i}{g}S\partial_i S^{-1}\;,\label{ansatzbegin1}\\
    \psi_A &= h_{AB} ST_AS^{-1}\;,\label{ansatzbegin2}\\
    S &= e^{i\varphi\beta\cdot T} \label{S} \makebox[.3in]{,} \beta\cdot T \equiv \beta|_q T_q  \;,
\end{align}
where $T_q$,  $q=1, \dots, N-1 $, are Cartan generators. 
\end{subequations}
Notice that $S$ is ill-defined along the $z$-axis, while $A_i$ must be smooth. 
Furthermore, $A_i$ should be a pure gauge when $\rho\rightarrow\infty$, so that the magnetic energy per unit length stored in the vortex is finite. 
Both issues can be resolved by defining
\begin{equation}
 \mathcal{A}_i = (a-1)\partial_i\varphi \, \beta\cdot T\;,
\end{equation}
with the boundary and regularity conditions
\begin{equation}
   a(\rho\rightarrow\infty)=1 \makebox[.5in]{,}   a(\rho\rightarrow0)=0\;.
\end{equation}
Clearly, the Higgs profiles must obey
\begin{equation}
\label{boundaryh}
    h_{AB}(\rho\rightarrow\infty)=v\delta_{AB}    
\end{equation}
so that their contribution to the energy per unit length is also finite.
The vortex charge is represented by $\beta = 2N \omega$, where $\omega$ is a weight of $\mathfrak{su}(N)$ 
and is closely connected with the 
$N$-ality $k$. For example, when $\omega$ is a weight of the fundamental representation, $\omega=\omega_1,\omega_2,...\omega_N$, then 
the vortex has $k=1$, while if it is a root $\alpha$, then the $N$-ality is that of the adjoint representation ($k=0$). 
A general $N$-ality can be reproduced by taking $\omega$ as the highest weight of the $k$-antisymmetric representation
\begin{equation}
\omega = \Lambda_k = \sum\limits_{i=1}^k \omega_i\;.
\end{equation}

Regarding the Higgs fields $\psi_A$ in Eq. (\ref{ansatzbegin2}), the number of profile functions $h_{AB}$ scales with $N^4$.
However, in the next section, we shall see that the vortex solutions display a collective behavior with a fixed 
reduced number of field profiles. A closer look at the local basis $n_A=ST_AS^{-1}$, 
\begin{subequations}
\label{localframe}
\begin{align}
    n_q &= T_q \;,\label{localcartan}\\
    n_\alpha &= \cos(\alpha\cdot\beta\, \varphi) \, T_\alpha - \sin(\alpha\cdot\beta\, \varphi) \, T_{\overline{\alpha}}\;,\label{localalpha}\\
    n_{\overline{\alpha}} &= \cos(\alpha\cdot\beta\, \varphi) \, T_\alpha + \sin(\alpha\cdot\beta\, \varphi) \,T_{\overline{\alpha}}\;,\label{localalpha2}
\end{align}
\end{subequations}
reveals that, whenever $\alpha\cdot\beta\ne0$, the elements $n_\alpha$ are ill-defined along the vortex line. On the other hand, 
the elements $n_q$ and $n_\alpha$ with $\alpha\cdot\beta=0$ have no defects. 
This leads to a natural splitting between $\psi_q$ and $\psi_\alpha,\psi_{\bar{\alpha}}$
\begin{subequations}
\label{ansatz2}
\begin{align}
    \psi_q&=h_{qp}T_q\;,\label{ansatz2.1}\\
    \psi_\alpha =\psi_{\bar{\alpha}}&= h_\alpha ST_\alpha S^{-1}\label{ansatz2.2}
\end{align}
\end{subequations}
and the regularity condition
\begin{equation}
\label{regularityh}
    h_\alpha(\rho\rightarrow0) = 0 \makebox[.5in]{\rm if }\alpha\cdot\beta\ne0\;.
\end{equation}

So far, the equations of motion read 
\begin{subequations}
\label{eq}
\begin{align}
    \frac{1}{\rho}\frac{\partial a}{\partial \rho} &- \frac{\partial ^2 a}{\partial \rho^2}=  g^2h_\alpha^2 (1-a)\left(\beta\cdot\gamma\right)\left(\gamma\cdot T\right)\;,\label{firsteq}\\
\nabla^2 h_{qp} &= \mu^2 h_{qp} + h_{\gamma}^2\kappa\gamma_q \gamma_p + \lambda h_{\gamma}^2 h_{ql}\gamma_l\gamma_p \;,\label{secondeq}\\
\nabla^2 h_{\alpha} &= (1-a)^2\left(\alpha \cdot \beta/\rho\right)^2h_{\alpha} +\mu^2 h_{\alpha}   \notag \\
&+ 2\kappa h_{\alpha} \alpha_q h_{qp} \alpha_p+ \kappa N_{\alpha,\gamma}^2 h_{\gamma}h_{\alpha+\gamma} + \lambda   h_{\alpha}^3 \alpha^2 \label{thirdeq} \\
&+\lambda h_{\gamma}^2h_\alpha N_{\alpha,\gamma}^2+\lambda h_\alpha \alpha_q h_{qp} h_{pl} \alpha_l \notag\;.
\end{align}
\end{subequations}
In Eq. (\ref{eq}), $\gamma$ is summed over all the roots except in Eq. (\ref{thirdeq}) where $\gamma \neq -\alpha$
and there is no summation over the repeated positive root $\alpha$. When $\gamma<0$, $h_\gamma = h_{-\gamma}$ is understood. Although smaller, 
the number of profiles in Eq. (\ref{eq}) still scales with $N^2$. In what follows, we shall further reduce their number 
by carefully studying the equations of motion. We shall initially  
address the simpler $k=1$ case and then we will extend the analysis to $k>1$.

\subsection{Case k=1}

In Ref. \cite{Oxman2013}, a reduced ansatz was constructed for $SU(2)$ and $SU(3)$, and it was numerically explored in 
Ref. \cite{Oxman:2016kjn}. Note that for $N \leq 3$ there is no 
variety in the possible string tensions as vortices with $k$ and $-k$ have the same tension, and for $N=3$ the $N$-ality $k=2$ is equivalent to $k=-1$.
In this subsection we shall extend the $k=1$ case for an arbitrary $N$, while the $k>1$ case will be worked out in the next subsection. In view of Eqs. (\ref{localframe}) and (\ref{thirdeq}), 
it is natural to propose a collective behavior that
only depends on the product $\alpha\cdot\beta$,
\begin{equation}
h_\alpha = h_{\overline{\alpha}} = \begin{cases} 
h_0\text{, if } \alpha\cdot\beta=0\;,\\
h\text{, if }\alpha\cdot\beta= 1\;. \\
\end{cases}
\end{equation}
As a consequence, Eq. (\ref{firsteq}) turns out to be
\begin{equation}
\label{profileeqa}
   \frac{1}{\rho}\frac{\partial a}{\partial \rho} - \frac{\partial ^2 a}{\partial \rho^2}=  g^2h^2 (1-a)\;. 
\end{equation}
 With regard to the Cartan sector,  Eq. (\ref{secondeq}) involves only three matrices: 
The $\rho$-dependent $\mathbb{H}\vert_{qp} = h_{qp}$ and the constant ones
\begin{equation}
\mathbb{A}\vert_{qp} = \sum\limits_{\alpha>0\;;\;\alpha\cdot\beta=1} \alpha\vert_q\alpha\vert_p\;;\;\mathbb{A}_0\vert_{qp} = \sum\limits_{\alpha>0\;;\;\alpha\cdot\beta=0} \alpha_0\vert_q\alpha_0\vert_p\;,
\end{equation}
which satisfy
\begin{subequations}
\label{A0}
\begin{align}
\mathbb{A}+\mathbb{A}_0 &= \frac{1}{2}\mathbb{I}\;,\label{A01}    \\
\mathbb{A}_0^2 &= \frac{N-1}{2N} \, \mathbb{A}_0\;\label{A02}.
\end{align}
\end{subequations}
Thus, we can use Eq. (\ref{A01}) to eliminate $\mathbb{A}$ and cast Eq. (\ref{secondeq}) into the form
\begin{equation}
\label{matrixH}
\big[ \big(\nabla^2-\mu^2-\frac{\lambda}{2}h^2\big)\, \mathbb{I} -\lambda(h_0^2-h^2)\, \mathbb{A}_0 \big] \mathbb{H} = \frac{\kappa}{2}h\mathbb{I} + \kappa(h_0-h)\mathbb{A}_0 \;.
\end{equation}
As the Laplacian is a scalar operator, the inversion of the matrix operator in the first member will 
be a power series in $\mathbb{A}_0$. Then, because of Eq. (\ref{A02}), the solution for $\mathbb{H}$ in Eq. (\ref{matrixH}) 
must be a linear combination of $\mathbb{I}$ and $\mathbb{A}_0$. We can define a pair of projectors, 
$\mathbb{M}_1 + \mathbb{M}_2 = \mathbb{I}$, $\mathbb{M}_i \mathbb{M}_j = \delta_{ij} \mathbb{I} $, by taking
\begin{equation}
\mathbb{M}_2 = \frac{2N}{N-1} \, \mathbb{A}_0  \;,
\end{equation} 
and write
\begin{equation}
\label{finalH}
\mathbb{H} = h_1\mathbb{M}_1+h_2\mathbb{M}_2\;.
\end{equation}
In this manner, if these profiles satisfy
\begin{subequations}
\begin{align}
\nabla^2 h_1 &= \mu^2 h_1 + (\kappa+\lambda h_1)h^2\;,\label{profileeqh1}\\
\nabla^2 h_2 &= \mu^2h_2 + \frac{h^2+(N-1)h_0^2}{N}(\kappa+\lambda h_2)\;,
\end{align}
\end{subequations} 
then the equations in the Cartan sector close.
Now, to simplify those for $h$ and $h_0$, we note that according to our conventions the coefficients 
$N_{\alpha,\gamma}$ are given by (see section 5.5 in Ref. \cite{Gilmore})
\begin{equation}
\label{N}
N_{\alpha,\gamma}^2 = \frac{1}{2}\alpha\cdot \alpha = \frac{1}{2N}\;,
\end{equation}
when $\alpha+\gamma$ is a root, and they are zero otherwise. Thus, in order to perform the summation over $\gamma$ in Eq. (\ref{thirdeq}),
we have to count the number of terms for each profile combination. For a fixed $\alpha$, the multiplicities are summarized in table \ref{one}.
\begin{table}
\caption{}
\begin{center}
\begin{tabular}{| c | c |}
\hline
Profile types & Number of terms\\
\hline
$(h_\alpha,\; h_\gamma,\; h_{\alpha+\gamma})=(h,h,h_0)$ & $N-2$ \\
\hline
$(h_\alpha,\; h_\gamma,\; h_{\alpha+\gamma})=(h,h_0,h)$ & $N-2$ \\
\hline
$(h_\alpha,\; h_\gamma,\; h_{\alpha+\gamma})=(h_0,h_0,h_0)$ & $2(N-3)$ \\
\hline
$(h_\alpha,\; h_\gamma,\; h_{\alpha+\gamma})=(h_0,h,h)$ & $2$\\
\hline
\end{tabular}
\end{center}
\label{one}
\end{table}
Combining these ingredients, the remaining Higgs equations can be simplified to
\begin{subequations}
\begin{align}
\nabla^2 h_0 &= \mu^2 h_0+ \frac{h_0}{N}(2\kappa h_2 + \lambda h_2^2+\lambda h_0^2) \\
&+ \frac{(\kappa+\lambda h_0)}{N}(h^2+(N-3)h_0^2)\;,\notag\\
\nabla^2 h &= \mu^2 h + \frac{(1-a)^2}{\rho^2}h+ \frac{\lambda}{2}h^3\notag\\ 
&+\frac{(N-2)}{2N}hh_0(2\kappa + \lambda h_0)+ \frac{(2\kappa+\lambda h_1)}{2(N-1)}hh_1 \\
&+ \frac{(N-2)}{2N(N-1)}(2\kappa + \lambda h_2)hh_2\;.\notag
\end{align}
\end{subequations}
They must be solved with the Higgs profiles approaching the vacuum value $v$ when $\rho\rightarrow\infty$, so as to 
comply with Eq. (\ref{boundaryh}), while $h(\rho)$ must also obey the regularity condition (\ref{regularityh}).

\subsection{Case $k>1$}

To solve the case $k >1$, we consider a general $\beta = 2N \Lambda^k$ in Eq. (\ref{S}). The reasoning to be followed is very similar to 
the previous one. The main difference is that we have to split the positive roots with $\alpha \cdot \beta =0$ into two categories: 
\begin{subequations}
\label{alphak}
\begin{align}
\tilde{\alpha}_0 &= \omega_{i\le k}-\omega_{j\le k}\;,\\
\alpha_0&= \omega_{i>k}-\omega_{j>k}\;.
\end{align}
\end{subequations}
The point is that $\alpha_0$ and $\tilde{\alpha}_0$ have a slightly different behavior. 
For example, there are $k(k-1)$ roots of type $\tilde{\alpha}_0$ and $(N-k)(N-k-1)$ roots of type $\alpha_0$, which 
generates a difference when counting the terms in (\ref{thirdeq}). Note that for $k=1$ there are no roots of type $\tilde{\alpha}_0$.  
The roots associated with a rotating $n_\alpha$ are given by
\begin{equation}
\alpha = \omega_{i\le k}-\omega_{j>k}\;.\\
\end{equation}
Thus, we are led to introduce three profiles in the $\alpha$-sector,
\begin{equation}
h_\alpha = \begin{cases}
\tilde{h}_0,\text{ if } \alpha=\tilde{\alpha}_0\\
h_0,\text{ if } \alpha=\alpha_0 \\
h,\text{ if }\alpha\cdot\beta = 1  \;.
\end{cases}
\end{equation}
In any case, the equation for $a$ remains that in (\ref{profileeqa}).
 This time, in order to solve the matrix part of Eq. (\ref{secondeq}) we use three matrices 
$\mathbb{I}$, $\tilde{\mathbb{A}}_0$ and $\mathbb{A}_0$ instead of two. 
Following a similar reasoning, we can introduce three projectors, $\mathbb{M}_1 + \mathbb{M}_2 + \mathbb{M}_3
= \mathbb{I}$, determined by
\begin{gather}
\label{matricesM}
\mathbb{M}_2=\frac{2N}{N-k}\mathbb{A}_0 \makebox[.5in]{,} \mathbb{M}_3=\frac{2N}{k}\mathbb{\tilde{A}}_0\;.
\end{gather}
In terms of them, the solution for $\mathbb{H}$ is
\begin{gather}
\mathbb{H} = h_1\mathbb{M}_1+h_2\mathbb{M}_2+h_3\mathbb{M}_3
\end{gather}
where $h_1$ satisfies Eq. (\ref{profileeqh1}), while $h_2$ and $h_3$ are determined by
\begin{subequations}
\begin{align}
\nabla^2h_2 &= \mu^2 h_2 + \left(\frac{k h^2+(N-k)h_0^2}{N}\right)(\kappa+\lambda h_2)\;,\label{profileseqh2}\\
\nabla^2h_3 &= \mu^2 h_3 + \left(\frac{(N-k)h^2+k\tilde{h}_0^2}{N}\right)(\kappa+\lambda h_3)\;.\label{profileseqh3}
\end{align}
\end{subequations}
Here, we begin to see how the center symmetry is made explicit by the ansatz. 
When the $Z(N)$ charge is changed from $k$ to $ N-k$, the equations for $h_2$ and $h_3$ get interchanged,
provided that $h_0$ and $\tilde{h}_0$ are also interchanged, which will be justified in the following discussion.

For a fixed $\alpha$, the mutliplicity of terms in Eq. (\ref{thirdeq}) with a given profile combination 
$(h_\alpha,\; h_\gamma,\; h_{\alpha+\gamma})$ 
are displayed in table \ref{two}.
\begin{table}
\caption{}
\begin{center}
\begin{tabular}{|c|c|c|c|}
\hline
Profile types & \# terms & Profile types & \# terms\\\hline
 $(h,h,\tilde{h}_0)$ & $(k-1)$  & $(\tilde{h}_0,h,h)$ & $2(N-k)$\\\hline
 $(h,h,h_0)$ & $(N-k-1)$ &$(\tilde{h}_0,\tilde{h}_0,\tilde{h}_0)$ & $2(k-2)$\\\hline
$(h,\tilde{h}_0,h)$ & $(k-1)$ &$(h_0,h,h)$ & $2k$\\\hline
 $(h,h_0,h)$ & $(N-k-1)$ & $(h_0,h_0,h_0)$ & $2(N-k-2)$\\\hline
\end{tabular}
\end{center}
\label{two}
\end{table}
In addition, in expressions such as the energy, where a sum over $\alpha$ is required, the above numbers should be multiplied by 
$k(N-k)$ if $n_\alpha$ rotates, by $\frac{k(k-1)}{2}$ if the root is of type $\tilde{\alpha}_0$, and by $\frac{(N-k)(N-k-1)}{2}$ 
if it is of type $\alpha_0$. With this information at hand, the equations for $h$, $h_0$ and $\tilde{h}_0$ become
\begin{subequations}
\label{profileseq}
\begin{align}
\nabla^2 h_0 &= \mu^2 h_0 + \frac{h_0}{N}(2\kappa h_2 + \lambda h_2^2 + \lambda h_0^2) \label{profileseqh0}\\
&+ \frac{(\kappa+\lambda h_0)}{N}(k h^2 + (N-k-2)h_0^2)\;,\notag\\
\nabla^2 \tilde{h}_0 &= \mu^2 \tilde{h}_0 + \frac{\tilde{h}_0}{N}(2\kappa h_3 + \lambda h_3^2+ \lambda \tilde{h}_0^2)  \label{profileseqth0}\\
&+ \frac{(\kappa+\lambda \tilde{h}_0)}{N}((N-k) h^2 + (k-2)\tilde{h}_0^2)\;,\notag\\
\nabla^2 h &= \mu^2 h +\frac{(1-a)^2}{\rho^2}h +\frac{\lambda}{2}h^3+ \frac{hh_1}{2k(N-k)}(2\kappa+\lambda h_1)\notag\\
&+\frac{N-k-1}{2N(N-k)}(2\kappa+\lambda h_2)hh_2+\frac{k-1}{2Nk}(\kappa+\lambda h_3)hh_3 \notag\\
&+ \frac{N-k-1}{2N}(2\kappa+\lambda h_0)hh_0+ \frac{k-1}{2N}(2\kappa+\lambda \tilde{h}_0)h\tilde{h}_0\;.\label{profileseqh}
\end{align}
\end{subequations}
with boundary conditions similar to those for $k=1$, where $h$ is the only profile with a regularity condition along the vortex line.
As anticipated, under $  k\rightarrow N-k$ we have
\begin{equation}
 h_2\leftrightarrow h_3 \makebox[.5in]{,} h_0\leftrightarrow\tilde{h}_0\;.
\end{equation}
Indeed, due to these properties, the center symmetry is made explicit: the energy of a vortex with charge $k$ and an
antivortex with charge $N-k$ are the same. 
Incidentally, it is easy to see that 
the differences $\Delta h = h_0-h_2$ and $\Delta \tilde{h} = \tilde{h}_0-h_3$ are governed by
\begin{subequations}
\begin{align}
    (\nabla^2-\mu^2)\Delta h&= \frac{\lambda h^2 + \lambda(N-k-1)h_0^2-\kappa h_0}{N}\Delta h\;,\\
(\nabla^2-\mu^2)\Delta \tilde{h}&=\frac{\lambda h^2 + \lambda(k-1)\tilde{h}_0^2-\kappa \tilde{h}_0}{N}\Delta \tilde{h}\;.
\end{align}
\end{subequations}
for which $h_0=h_2$ and $\tilde{h}_0=h_3$ are solutions. This obviously holds for $k=1$ and leads to a welcomed additional reduction 
in the number of profiles.

Replacing the ansatz in the energy functional for the action (\ref{action}), we find 
\begin{eqnarray}
E &=& \int d^3x\, \frac{k(N-k)}{\rho^2}\left(\frac{|\nabla a|^2}{g^2}+ h^2(1-a)^2\right)\notag\\
&&+\frac{1}{2}|\nabla h_1|^2+\frac{1}{2}\mu^2h_1^2+\frac{(N-k)^2-1}{2}(|\nabla h_2|^2+\mu^2h_2^2)\notag\\
&&+\frac{k^2-1}{2}(|\nabla h_3|^2+\mu^2h_3^2)+k(N-k)(|\nabla h|^2+\mu^2h^2)\notag\\
&&+\lambda \frac{k(N-k)}{4}h^4+C_1 h^2 + C_2\;,
\end{eqnarray}
where $C_1$ and $C_2$ are given by
\begin{eqnarray}
C_1 &=& \frac{h_1}{2}(2\kappa+\lambda h_1)+\frac{k(N-k)^2-k}{2N}(2\kappa+\lambda h_2)h_2\notag\\
&&+\frac{(N-k)(k^2-1)}{2N}(2\kappa+\lambda h_3)h_3\;,\notag\\
C_2 &=& \frac{(N-k)^3+k-N}{N}\left(\kappa \frac{h_2^3}{3}+ \lambda\frac{h_2^4}{4}\right)+\kappa\frac{k^3-k}{3N}h_3^3\notag\\
&&\lambda\frac{k^3-k}{4N}h_3^4-(d^2-1)\left(\frac{\mu^2v^2}{2}+\frac{\kappa v^3}{3}+\frac{\lambda v^4}{4}\right) \;.\notag
\end{eqnarray}

A particularly interesting region in parameter space is $\mu^2=0$. 
In this case, except for $a$ and $h$, the profiles are frozen at the vacuum value $v$. This is possible because only $a$ and $h$ satisfy 
regularity conditions at $\rho=0$. Moreover, on the vortex ansatz, the nontrivial Higgs profiles  
$a$ and $h$ get Abelianized in the sense that they satisfy the usual Nielsen-Olesen (NO) equations. 
This is interesting because the YM chromoelectric field distribution obtained from the lattice 
is precisely that of the NO vortex-string \cite{Cosmai2017}.
The crucial difference is that in our case $N$-ality is automatically implemented due to the
underlying non-Abelian structure. 
Furthermore, at $\mu^2=0$, a direct calculation shows that the collective behavior gives rise to 
an exact Casimir scaling of the energy per unit vortex length (string tension)
\begin{equation}
\label{oursigma}
\sigma_k = k(N-k)\, \sigma_{\text{NO}}\;.
\end{equation}
Indeed, apart from a factor $(N+1)^{-1}$, the factor $k(N-k)$ is precisely the quadratic Casimir of the k-antisymmetric representation. 
In other words, an exact Casimir law
\begin{equation}
\label{Casimirlaw}
\sigma_k = \frac{C_2(A_k)}{C_2(F)} \, \sigma_1\;
\end{equation}
is analytically verified at $\mu^2=0$.

\section{\label{Numerics}Numerical solutions}

In principle, the numerical exploration of the model parameter space $(g,\mu^2,\kappa,\lambda)$ is a hard task since it is four-dimensional. Fortunately, we can reduce it to two dimensions by a simple rescaling, defining the dimensionless quantities
\begin{eqnarray}
\bar{x}_i = -\frac{\kappa}{g}x_i \makebox[.2in]{,} \bar{g}=1 \makebox[.2in]{,}\bar{\mu} = -\frac{g}{\kappa} \mu 
\makebox[.2in]{,}\bar{\kappa}=-1\,,\,\bar{\lambda}=\frac{\lambda}{g^2} \;, \nonumber
\end{eqnarray}
which implies the energy per unit lengh rescaled as
\begin{equation}
\label{Erescaling}
    \sigma(g,\mu^2,\kappa,\lambda) = -\frac{\kappa}{g^3} \, \sigma(1,\bar{\mu^2},-1,\bar{\lambda})\;.
\end{equation}
Then, for a given $N$-ality $k$, the ratio $\frac{\sigma_k}{\sigma_1}$ can only depend on $\bar{\mu}^2$, $\bar{\lambda}$. 
 Furthermore, when computing the string tension ratios, we observed that they essentially depend on the combination $\bar{\mu}^2 \bar{\lambda}$, 
 so we will also fix $\bar{\lambda}=1$ when evaluating this ratio. It is 
important to underline that the reduction from four parameters to one applies only to $\frac{\sigma_k}{\sigma_1}$ while
other observables may display a more complex behavior. For example, another important quantity we can always fit is the fundamental 
string tension $\sigma_1$. For every $\bar{\mu}^2$ and $\bar{\lambda}$, including $\bar{\lambda}\ne 1$, we  can evaluate the rescaled string 
tension and then set the proper $\kappa$ and $g$ in Eq. (\ref{Erescaling}) 
to obtain the well-established value $\sigma_1=(440$\, Mev$)^2$.
With regard to the numerical procedure, we initially discretized the coupled equations for $a,\,h,\,h_1,\,h_2$ and $h_3$. For this aim, 
we used finite differences with a range $\bar{\rho}\in[0.001,10]$ partitioned into 150 points. Then, we randomly swept over the domain updating each site using the relaxation method until the desired degree of convergence was met. All the simulations were implemented in \textsc{Mathematica}.
 We defined an error function as the modulus of the deviations summed over the various equations and
integrated over the domain,  using it to establish a numerical convergence criterion. 

\begin{figure}[h]
  \centering
    \includegraphics[width=0.45\textwidth]{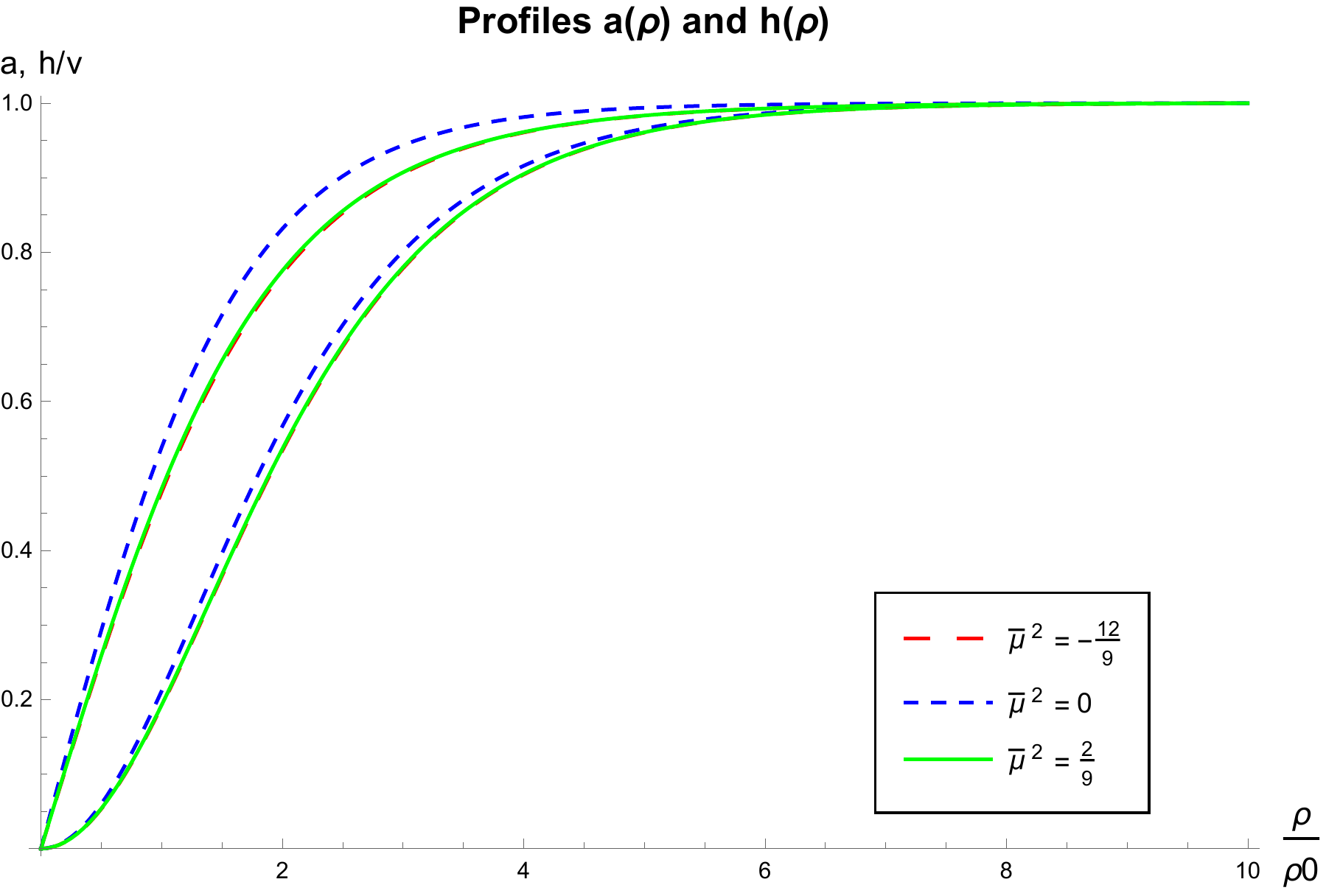}
  \caption{Profiles $a(\rho)$ and $h(\rho)$ for various $\bar{\mu}^2$. The profile $a$ is that with a linear behavior around $\rho=0$.}
 \label{ProfilePlotsah}
  \end{figure}

\begin{figure}[h]
  \centering
    \includegraphics[width=0.45\textwidth]{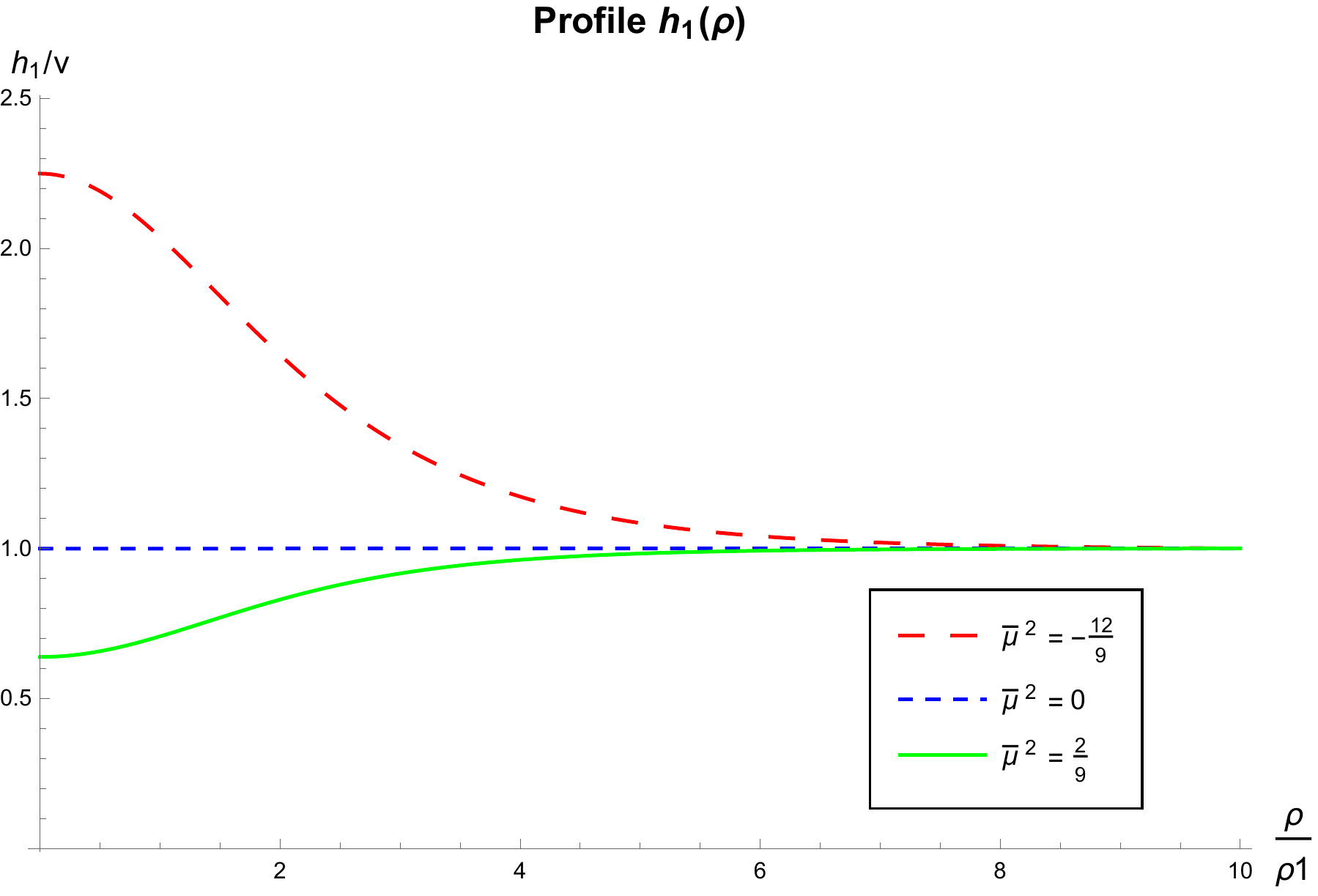}
  \caption{Profile $h_1(\rho)$ for various $\bar{\mu}^2$.}
  \label{ProfilePloth1}
  \end{figure}

\begin{figure}[h]
  \centering
    \includegraphics[width=0.4
    \textwidth]{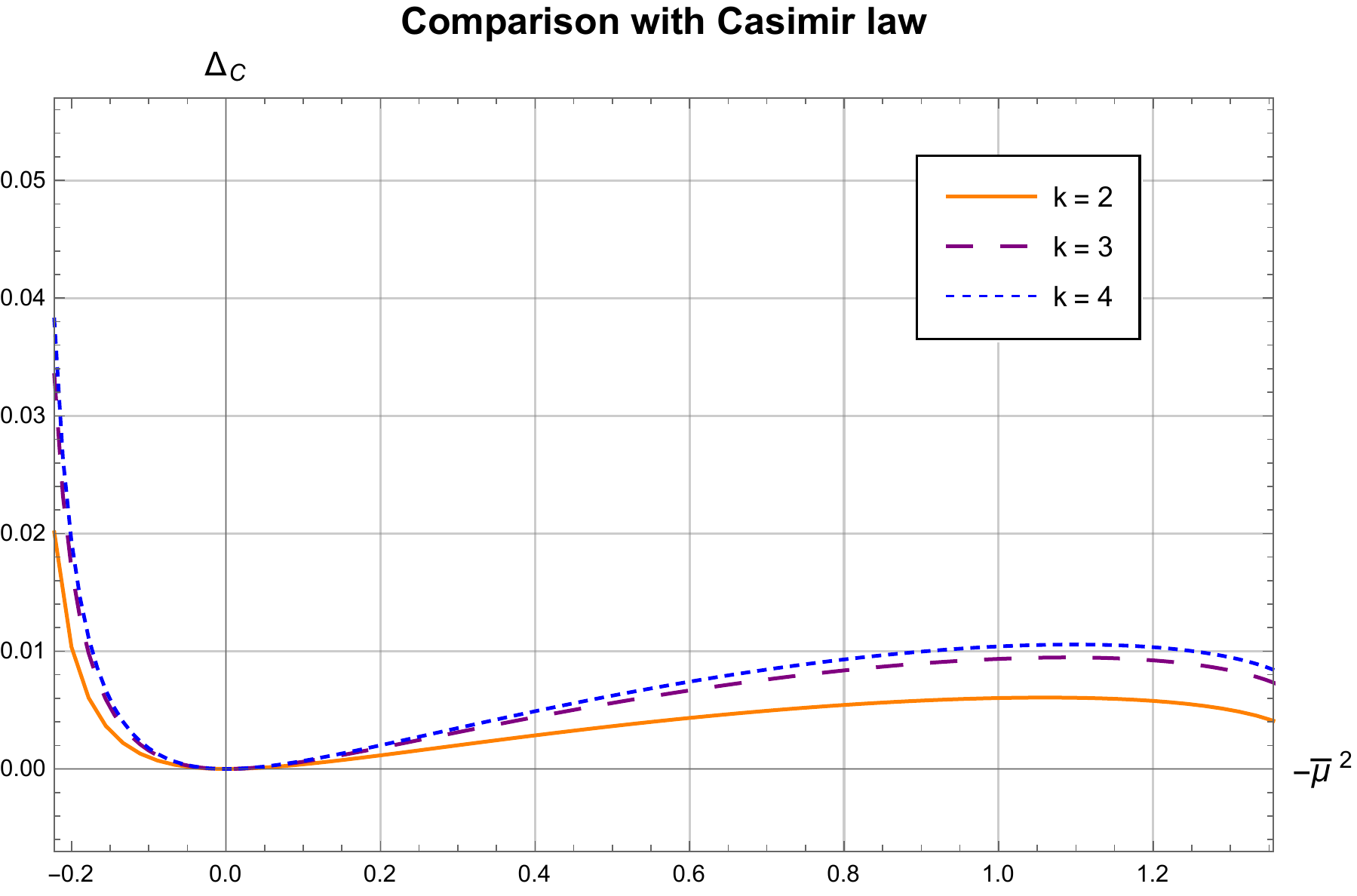}
  \caption{Plot of $\Delta_C(k)$ with $N=8$. Notice the region depicted is that where the SSB takes place, including positive $\bar{\mu}^2$.}
  \label{CasimirLawPlot}
  \end{figure}

\begin{figure}[h]
  \centering
    \includegraphics[width=0.4\textwidth]{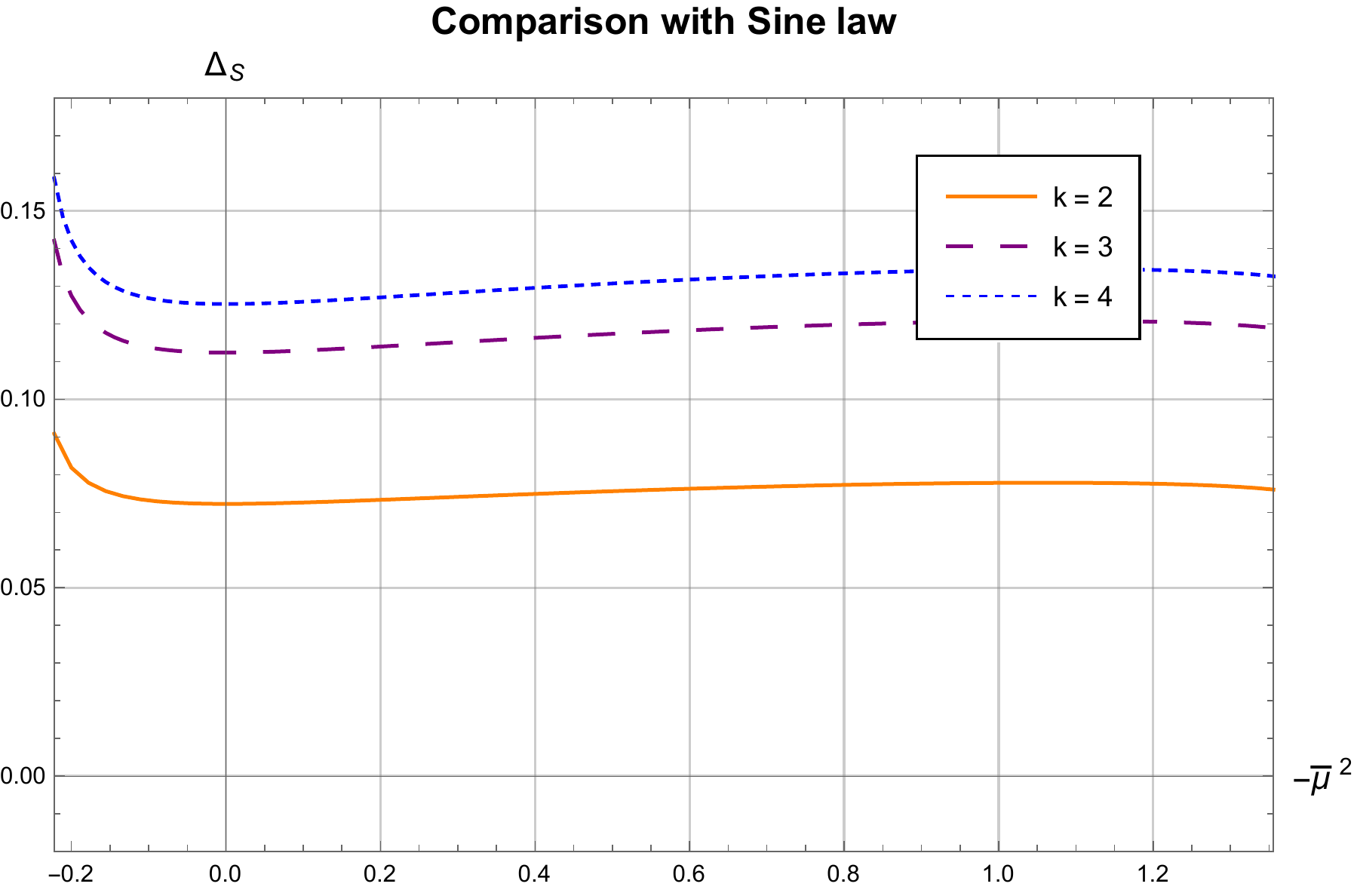}
  \caption{Plot of $\Delta_S(k)$ with $N=8$. The deviations are much larger in the whole region explored.}
  \label{SineLawPlot}
  \end{figure}

In Fig. \ref{ProfilePlotsah}, we plot $a(\rho)$ and $h(\rho)$ for various values of $\mu^2$, all of them with $g=\lambda=-\kappa= 1$.
Note that there are only small changes in the whole range considered. Since this seems to be true for other values of $g$, $\kappa$ and $\lambda$, 
we expect these profiles to be well approximated by those of the Nielsen-Olesen vortex. 
On the other hand, Fig. \ref{ProfilePloth1} shows that the profile $h_1$ is more influenced by changes in $\mu^2$. 
A similar behavior was also observed for $h_2$ and $h_3$.
In Fig. \ref{CasimirLawPlot}, we plot the quantity
\begin{equation}
    \Delta_C(k) = 1- \frac{N-1}{k(N-k)}\frac{\sigma_k}{\sigma_1} \;,
\end{equation} 
for $N=8$ and various values of $k$. It measures deviations between the Casimir law. At $\bar{\mu}^2=0$, this function passes by zero, a point where we 
showed an exact Casimir scaling. The simulations did not converge well for $\bar{\mu}^2 < -\frac{12}{9\bar{\lambda}}$.
It is interesting to note that the Casimir law is only slightly deviated from in the whole region we were able to explore.
In addition, as $\Delta_C(k)$ is positive, the scaling law of the model is slightly below the Casimir law. Recalling 
that the Sine law lies above the Casimir, it is not a surprise that in the whole range the model shows larger deviations
when compared with the Sine law (cf. Fig. \ref{SineLawPlot}), via the relative difference 
\begin{equation}
    \Delta_S(k) =  1- 
    \frac{\sin\left(\frac{\pi}{N}\right)}{\sin\left(\frac{k\pi}{N}\right)}\frac{\sigma_k}{\sigma_1} \;.
\end{equation}

\section{ \label{Conclusions} Conclusions}

Considering a dual YMH effective model with $N^2-1$ adjoint Higgs fields, we were able to develop an ansatz for a topologically stable static vortex carrying charge in the $k$-antisymmetric representation. The model has four parameters: the gauge coupling constant $g$, 
plus the quadratic ($\mu^2$), cubic ($\kappa$), and quartic ($\lambda$) couplings in the Higgs potential. 
We focused in the region $\mu^2<\frac{2}{9}\frac{\kappa^2}{\lambda}$, where the 
$SU(N)_{\rm color}$ is spontaneously broken to $Z(N)$ and the vacuum manifold is 
given by ${\rm Ad}(SU(N)) = SU(N)/Z(N)$, thus implementing $N$-ality. By using the algebraic structure of the model, especially that concerning the weights and roots of $SU(N)$, we showed that a collective behavior takes place. For $k=1$, the many adjoint scalar field equations are closed in terms of the 
profiles $h$, $h_1$ and $h_2$, while for $k>1$ only an additional profile $h_3$ is required. Since
this is valid for every value of $N$ and $k$, it allows
for a simple numerical simulation. Furthermore, when $\mu^2=0$, we found an exact Casimir law and nontrivial profiles coinciding with those of the Nielsen-Olesen vortex. 
This is compatible with the observed string tension and in agreement with the chromoelectric field distribution obtained in the lattice.
Finally, upon an appropriate rescaling, the dependence of string tension ratios on the model parameters was reduced from four to two adimensional quantities: $\bar{\mu}^2$ and $\bar{\lambda}$. 
 This made it easier to numerically explore the parameter space by using the relaxation method. We noticed that the scaling law depends in fact on the particular combination $\bar{\mu}^2\bar{\lambda}$ and that it is very stable throughout the parameter space. In particular, 
taking $N=8$ as an example, we observed that it deviates by at most $4\%$ from the exact Casimir law at $\bar{\mu}^2=0$. 

Our analysis encourages a thorough exploration of the interplay between ensembles observed in pure Yang-Mills lattice simulations, 
the associated large distance effective field description, the implied asymptotic properties, and their comparison with Monte Carlo calculations. 
Some of these connections were successfully verified in the model analyzed here. 

\section*{Acknowledgements}

The Conselho Nacional de Desenvolvimento Cient\'{\i}fico e Tecnol\'{o}gico (CNPq), the Coordena\c c\~ao de Aperfei\c coamento de 
Pessoal de N\'{\i}vel Superior (CAPES), and the Funda\c c\~{a}o de Amparo \`{a} Pesquisa do Estado do Rio de Janeiro (FAPERJ) are 
acknowledged for their financial support.

\end{document}